# Constraining neutron-star equation of state using heavy-ion collisions

C.Y. Tsang (曾浚源), M.B. Tsang (曾敏兒)[#], P. Danielewicz, and W.G. Lynch (連致標)
*National Superconducting Cyclotron Laboratory and the Department of Physics and Astronomy*
*Michigan State University, East Lansing, MI 48824 USA*
and
F.J. Fattoyev
*Center for Exploration of Energy and Matter*
*Department of Physics, Indiana University, Bloomington, IN 47408 USA*

Abstract

The LIGO-Virgo collaboration's ground-breaking detection of the binary neutron-star merger event, GW170817, has intensified efforts towards the understanding of the equation of state (EoS) of nuclear matter. In this letter, we compare directly the density-pressure constraint on the EoS obtained from a recent analysis of the neutron-star merger event to density-pressure constraints obtained from nuclear physics experiments. To relate constraints from nuclear physics to the radii and the tidal deformabilities of neutron stars, we use a large collection of Skyrme density functionals that describe properties of nuclei to calculate properties of $1.4 M_\odot$ neutron stars. We find that restricting this set of Skyrme equations of state to density functionals that describe nuclear masses, isobaric analog states, and low energy nuclear reactions does not sufficiently restrict the predicted neutron-star radii and the tidal deformabilities. Including pressure constraints on the EoS around twice saturation density ($2 \times 2.74 \times 10^{14}$ g/cm$^3$), obtained from high energy nucleus-nucleus collisions, does constrain predicted radii and tidal deformabilities to be consistent with the results obtained from the analysis of GW170817. We discuss how new measurements of nucleus-nucleus collisions can improve these constraints on the EoS to be more restrictive than the current constraints from the GW170817 merger event.

Email addresses:

[#]Corresponding Author: tsang@nscl.msu.edu
C.Y. Tsang: tsangc@nscl.msu.edu
W.G. Lynch: lynch@nscl.msu.edu
P. Danielewicz: danielewicz@nscl.msu.edu
F.J. Fattoyev: ffattoye@indiana.edu

The equation of state (EoS) of nuclear matter relates temperature, pressure and density of a nuclear system. It governs not only properties of nuclei and neutron stars but also the dynamics of nucleus-nucleus collisions and that of neutron-star mergers. The amount of ejected matter from the merger, which subsequently undergoes nucleosynthesis to form heavy elements up to Uranium and beyond [1-3] depends on the EoS. So does the fate of the neutron-star merger including; whether the colliding neutron stars collapse promptly into a black hole, remain a single neutron star, or form a transient neutron star that collapses later into a black hole [4]. The recent observation of a neutron-star merger event, GW170817, provides insight into the properties of nuclear matter and its equation of state (EoS) [5-8]. In this letter we examine the consistency of the EoS constraints obtained from laboratory experiments and GW170817 as shown in Fig. 1. We also assess the prospects for more stringent comparison between astrophysical and laboratory measurements.

During the inspiral phase of a neutron-star merger, the gravitational field of each neutron star induces a tidal deformation in the other [9]. The influence of the EoS of neutron stars on the gravitational wave signal during inspiral is contained in the dimensionless quantity tidal deformability also known as tidal polarizability, $\Lambda = \frac{2}{3} k_2 \left(\frac{c^2 R}{GM}\right)^5$, where $G$ is the gravitational constant and $k_2$, is the dimensionless Love number [5, 9], $R$ and $M$ are the mass and radius of a neutron star. $k_2$, is sensitive to the compactness parameter ($M/R$) where $M$ is the mass of the neutron star. As the knowledge of the mass-radius relation uniquely determines the neutron-star matter EoS [10-13], both $k_2$ and $R$ depend on the EoS.

In the original GW170817 analysis of late-stage inspiral [5], an upper limit of $\Lambda < 800$ is obtained for a $1.4 M_\odot$ neutron star under low spin scenario. The mass-weighted $\Lambda$ have the same values. In a recent analysis, the mass-weighted $\Lambda$ values are updated to $300^{+420}_{-230}$ [7]. By further assuming both neutron stars have the same EoS, more restrictive $\Lambda$ values of $190^{+390}_{-120}$ and $R$ values of $11.9^{+1.4}_{-1.4}$ km were obtained [8]. The latter analysis also extracted the pressure as a function of density, plotted as the blue hatched area in Figure 1.

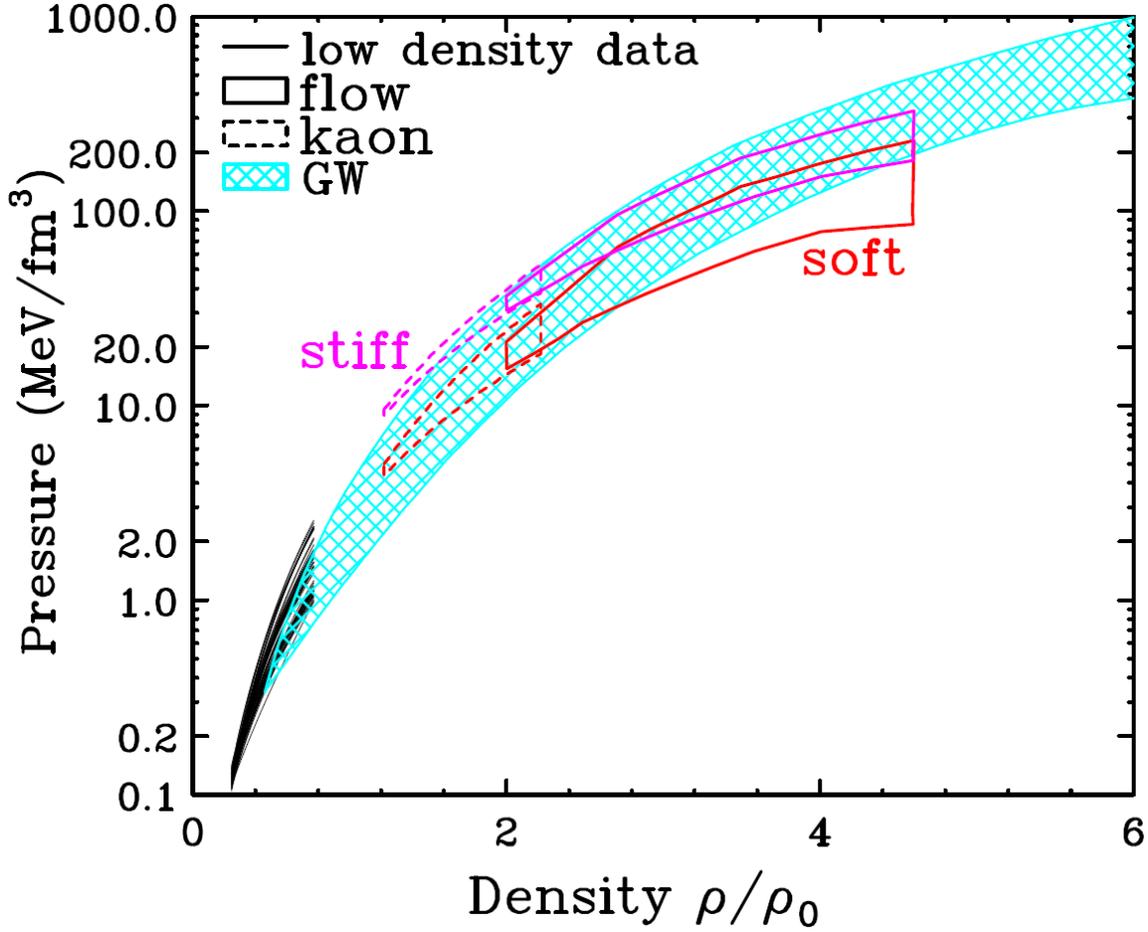

**Fig. 1:** Experimental and astrophysical constraints on equation of state in pressure vs. density. The hatched region represents the GW constraint [8], after converting the unit for pressure to MeV/fm$^3$ and the density to units of saturation density, $\rho_0$=2.74×10$^{14}$ g/cm$^3$, to allow direct comparisons to nuclear physics constraints. Solid and dashed contours display constraints from flow measurements [37] and kaon measurements [38, 39], respectively. The upper (labelled as stiff) and lower (labelled as soft) constraints correspond to the addition of symmetry pressure from Eqs. (2) and (3), respectively, to the symmetric matter pressure. The black curves below 0.7$\rho_0$ represent Skyrme functionals satisfying the density dependent symmetry energy data from 0.25< $\rho/\rho_0$ < 0.7 in [29].

The nuclear EoS of cold homogenous matter can be specified in terms of the energy per nucleon of the hadronic system. Within the parabolic approximation, the EoS of cold nuclear matter can be divided into a symmetric matter contribution that is independent of the neutron-proton asymmetry and a symmetry energy term, proportional to the square of the asymmetry [14],

$$E(\rho, \delta) = E(\rho, \delta=0) + S(\rho)\delta^2 \qquad (1)$$

where the asymmetry is defined as $\delta = (\rho_n - \rho_p)/\rho$. Here, $\rho_n$, $\rho_p$ and $\rho = \rho_n + \rho_p$ are the neutron, proton and nucleon densities, and $S(\rho)$ is the density dependent symmetry energy. $E(\rho, \delta)$ can be modeled in terms of a Skyrme Hartree-Fock density functional that describes nuclear properties [15,16].

To connect nuclear physics observables to $1.4M_\odot$ neutron-star observables, we start with a large collection of Skyrme nuclear density functionals. We extrapolate the functionals to densities and asymmetries comparable to that for a neutron star as described below. We adopt the method described in Ref. [17] to calculate properties of a $1.4M_\odot$ neutron star.

Different forms of the EoS are used in four density regions defined as the outer crust, inner curst, outer core and inner core: 1) For the outer crust, consisting of a Coulomb lattice of neutron-rich nuclei embedded in a degenerate electron gas, we use the well-known EoS of Ref. [18] up to the density where neutron drip occurs. 2) In the inner crust, which consists a neutron gas in coexistence with nuclei or even nuclear pasta [19-23], we employ the EoS from [24], and connect to the outer core through a polytropic EoS of the form $P = A + K\left(\frac{E}{V}\right)^{\frac{4}{3}}$ [25, 26]. The parameters $A$ and $K$ are varied to match the crustal EoS to that of a liquid core at the crust-core transition density, $\rho_{TD}$, predicted by our choices for the Skyrme interactions. 3) In the region of $\rho_{TD} < \rho < 3\rho_0$, which corresponds to the outer core, we use various nuclear Skyrme density functionals, modified to take beta equilibrium between nuclear, electronic and muonic matter into account. 4.) For high density regions of $\rho > 3\rho_0$, a polytropic EoS of the form $K'\rho^\gamma$ is used to extend the EoS to the central density region of a neutron star. The constants $K'$ and $\gamma$ are fixed by the conditions that the pressure at thrice the normal nuclear density, $P(3\rho_0)$ matches the pressure from the Skyrme density functionals and that the polytrope pressure at $7\rho_0$ is such that the EoS can support a $2.0M_\odot$ neutron star. For the low mass neutron stars that are considered here, the density region above $3\rho_0$ does not affect the relevant properties of the neutron stars significantly.

Following this procedure, we arrived at 163 Skyrme interactions from our collection of 216 interactions [15] that can support a $2.0M_\odot$ neutron star. Each interaction, represented by an open circle in Fig. 2, makes a unique prediction for the neutron-star radius and tidal deformability. For reference, the red curve is a fitted relationship of $R$ and $\Lambda$ from a "generic" neutron-star EoS [27]. The common trend exhibited by the red curve and the open circles reflects the fact that $R$ and $\Lambda$ are correlated as described by Eq. (2) and that both are largely determined by the pressure at $2\rho_0$ [8, 28]. Our results are consistent with those from EoS based on relativistic mean-field interactions [10] represented by the open red squares using analogous methodology. Above $\Lambda > 600$, our calculations produce larger radii and deviate from the red curve. This may be due to inclusion of the crustal EoS at very low density in our calculations. The range of the updated values of $\Lambda=70\text{-}720$ obtained in [7] is represented by the larger light blue-hatched region. The blue rectangle indicates the more restricted region of $\Lambda=70\text{-}580$ and $R = 10.5\text{-}13.3\ km$ obtained in [8]. This tighter constraint is referred simply as "GW" constraints.

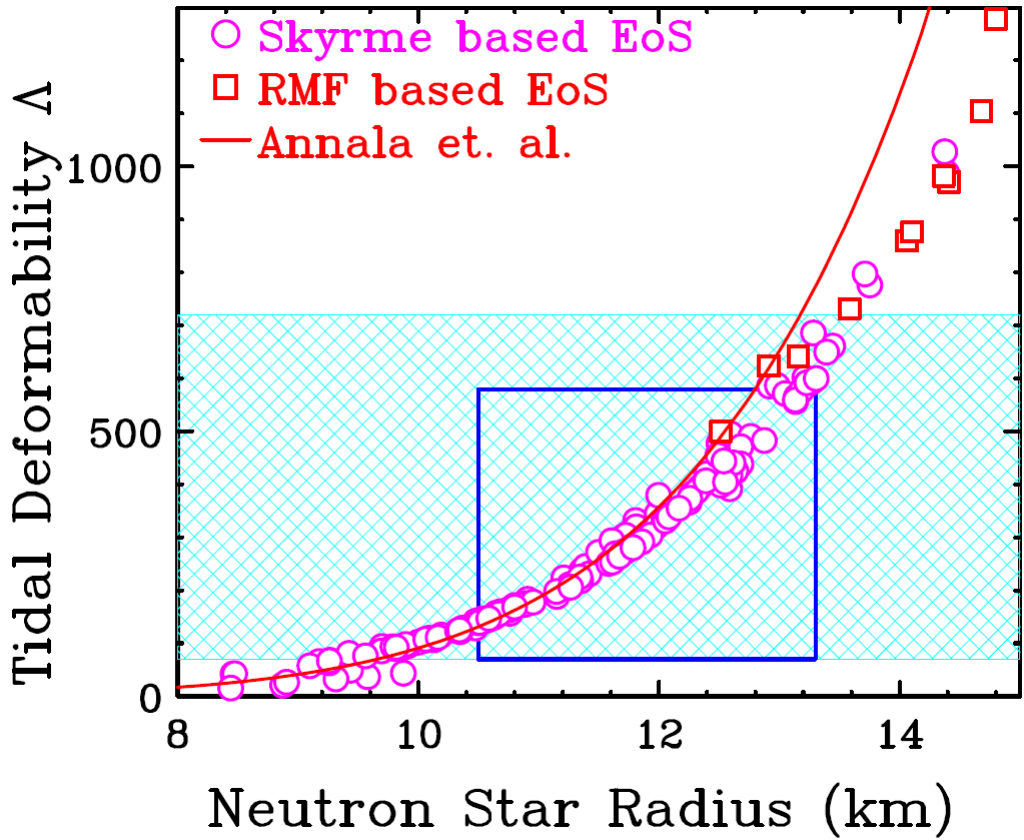

**Fig. 2:** Correlation between neutron-star tidal deformability and radii from current calculations (open circles) and from Ref. [10] (open squares). The curve is from Ref. [27]. The hatched areas represent constraints from recent GW170817 analysis [7,8].

In the past two decades, the nuclear EoS has been studied over a range of densities $0.25\rho_0 < \rho < 4.5\rho_0$ in nuclear structure and reaction experiments [29-41]. This density range is comparable to that found in neutron stars. However, in the latter context, the EoS must be extrapolated to environments where the density of neutrons greatly exceeds the density of protons. The extrapolation of the EoS to neutron–rich matter depends on $S(\rho)$ in Eq. (1).

At sub-saturation densities, $\rho < \rho_0$, information on EoS is obtained in the context of bound nuclei and via nuclear collisions [29-36]. By evaluating the chi-squares per degree of freedom, $c_u^2$, between experimental symmetry energy values at various densities extracted in Ref. [29], 69 Skyrme functionals that satisfy the low density data with $c_u^2 < 2$ are plotted as black curves in the lower left corner of Fig. 1. There is reasonable overlap between the GW constraint and the experimental constraints. Nuclear experiments, typically probe matter over a restricted range of densities. By varying energy and impact parameters of the heavy ion collisions (HIC), densities lower than that probed by the neutron-star merger observation, can be created in the laboratory.

Next we examine the constraints from heavy ion collisions that exist in the supra-saturation density region. Measurements of collective flow and kaon production in energetic nucleus-nucleus collisions have constrained the EoS for symmetric matter, $E_0(\rho, \delta=0)$, at densities up to 4.5 times saturation density [37-41]. In Ref [37], the symmetric matter constraints in pressure vs. density were determined from the measurements of transverse and elliptical flow from Au+Au collisions over a range of incident energies from 0.3 to 1.2 GeV/u. In Refs. [38, 39], a similar constraint from $1.2\rho_0$ to $2.2\rho_0$ was obtained from the kaon measurements.

To estimate the additional pressure coming from the symmetry energy, we use the softest and stiffest symmetry energy functionals proposed in Ref. [42] for utilization in neutron-star calculations:

$S(\rho)_{stiff}=12.7\ MeV\times(\rho/\rho_0)^{2/3}+38MeV\times(\rho/\rho_0)^2/(1+\rho/\rho_0)$ (2)

$S(\rho)_{soft}=12.7\ MeV\times(\rho/\rho_0)^{2/3}+19MeV\times(\rho/\rho_0)^{1/2}$ (3)

For convenience, we label the functionals in Eq. (2) and Eq. (3) as "stiff" and "soft", respectively. Adding the pressure from each of these two symmetry energy functionals to the pressure from the symmetric matter constraint results in two contours (solid lines) shown in Fig. 1 [37]. These heavy ion constraints were found to agree with the Bayesian analyses of the neutron-star mass-radius correlation in [43]. They also agree reasonably well within the pressure-density constraints extracted in [8] represented by the light blue hatched area in Fig. 1. At density $> 3\rho_0$, the GW constraint seems to prefer symmetry energy stiffer than that of Eq. (2).

Predictions for tidal deformability and radii using different nuclear models for the EoS are being actively pursued and the predicted $\Lambda$ values vary from ~50 to 1100 [10, 27, 44-49]. To illustrate the consequence of measurements from heavy ion collisions, we focus on the experimental and neutron-star merger observables at $0.67\rho_0$ and $2\rho_0$.

For the low density region, we use the accurate symmetry energy (~25 MeV) that have been derived from the analysis of nuclei masses [30, 31] and isobaric analog states [15], each of which probes the $S(r)$ at $r \approx \frac{2}{3}r_0$. We calculate both the symmetry energy at such density and the neutron-star properties of tidal deformability (left panels) and radius (right panels) in Fig. 3(a) and 3(b) (upper panels). For reference, we show, by red horizontal dashed lines, the symmetry constraints from double magic nuclear masses of Ref. [30], and, by the blue horizontal solid lines, the corresponding constraints obtained from a wider range of masses and isobaric analog states [15, 31]. In accordance with our selection criterion, the black open points lie within these low-density nuclear constraints. As the correlation between symmetry energy and $\Lambda$ or $R$ is very week, this selection criterion, alone, does not restrict the allowed values for $\Lambda$, which scatters over the range from 20 to 1000. Such lack of correlation between properties of nuclei and neutron-star properties at low density should be expected. The EoS at high density reflects dominant contributions from strongly repulsive three-body interactions and few-body correlation effects that are not adequately probed by masses, isobaric analog states and other observables at low density; therefore, they remain uncertain [43]. The role of the uncertainty in three-body interactions has been elucidated in Refs. [43, 50].

The density region of $2\rho_0$ has been identified to be most sensitive to the neutron-star radii [48]. At this density our calculations (open circles) show strong correlation between the pressure and the neutron-star properties such as tidal deformability and radii in Figs. 3(c) and 3(d), respectively. Similar correlation is also observed in [51]. The colored horizontal bars in the bottom panels of Fig. 3 represent the pressure expected at $2\rho_0$ after adding the symmetry pressure from Eq. (2) (upper green bars) or Eq. (3) (lower red bars). The upper bound of $P<40\ MeV/fm^{-3}$ from [8] excludes pressures much greater than those which arise from Eq. (2). For pressure consistent with a "stiff" symmetry energy, a lower bound in $\Lambda$ and $R$ could be provided by HIC results. On the other hand, a measurement consistent with the "soft" EoS (lower red bars), would provide both a lower and upper bound on $\Lambda$ and $R$ values. The uncertainty increases with the softness of the EoS, but it still could be a factor of two to three smaller than the uncertainty of the GW constraint. With additional events, one can hope that the GW constraints will improve. The expected improvement in the HIC constraints will allow us to understand what these combined constraints imply about the nature of strongly interacting matter.

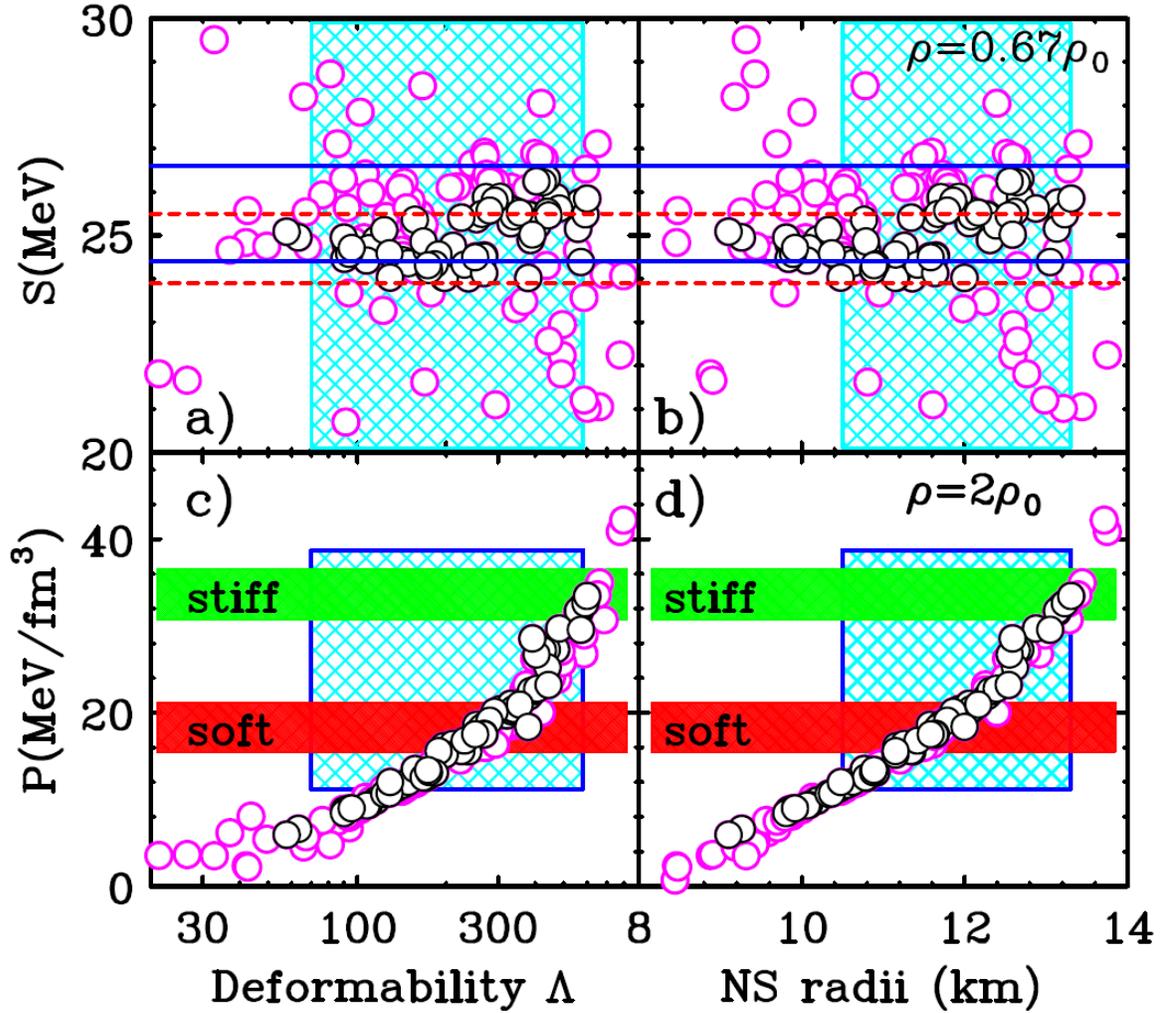

**Fig. 3:** Symmetry energy at $0.67\rho_0$ (upper panels) and pressure at $2\rho_0$ (lower panels) vs. neutron-star properties, tidal deformability (left panels) and radii (right panels). Black open circles correspond to interactions that agree with the low density data while the magenta circles represent calculations that do not. a.) & b.) Calculated symmetry energy, tidal deformability and radii from the Skyrme interactions at $0.67\rho_0$. The red dashed and solid blue horizontal lines are experimental symmetry energy values. c.) &d.) Calculated pressure, tidal deformability and radii from the Skyrme interactions at $2\rho_0$. The top (green) and bottom (red) horizontal bars represent the predicted experimental stiff and soft constraints, respectively.

In Fig. 1, the stiff and soft contours from flow experiments begin to overlap at high density $\rho>2.5\rho_0$, suggesting that the measurements aiming at extracting the symmetry pressure at very high densities could be associated with large uncertainties. The splitting between stiff and soft symmetry energy constraints increases with decreasing density. However, below the saturation density, the correlation between symmetry energy and neutron-star properties is weak [51]. Our calculations show that the correlation improves with density and a reasonably good correlation is achieved around $1.5\rho_0$. Thus the density region between $1.5\rho_0 – 2.5\rho_0$ currently represents a "sweet" spot for experimental exploration of the symmetry pressure for asymmetric nuclear matter. Higher density regions should also be explored especially if experimental uncertainties in symmetric matter can be reduced by new analysis or with better data [37,40].

In the case of the symmetric matter EoS [37], accurate constraints on the pressure require detailed calibration of "barometers" constructed from collective flow and particle production observables with the aid of transport model simulations [37,52]. The same algorithm will be required to extract pressure arise from symmetry energy. The evaluation of transport models needed to interpret the HIC data is currently under way [53, 54].

In summary, while symmetry energy data extracted at low density do not correlate strongly with neutron-star properties, heavy ion collision experiments testing twice the normal nuclear matter density may provide tighter constraints on the tidal deformability and thus the corresponding neutron-star radii. Current experimental plans anticipate constraints on the symmetry pressure that can distinguish the stiff and soft symmetry energy terms in the EoS [55, 56]. Our work suggests that it is feasible to expect experimental constraints obtained in this density region to be more stringent than the current GW analysis.

**Acknowledgement**:

We would like to thank Prof. C. J. Horowitz for advice and fruitful discussions. This work was partly supported by the US National Science Foundation under Grant PHY-1565546 and by the U.S. Department of Energy (Office of Science) under Grants DE-SC0014530, DE-NA0002923, DE-FG02-87ER40365 (Indiana University) and DE-SC0018083 (NUCLEI SciDAC-4 Collaboration). This work was stimulated by discussions with participants at the INT-JINA Symposium "First multi-messenger observations of a neutron-star merger and its implications for nuclear physics".

# References

[1] S.Rossweg, M. Liebendoerfer, F.-K.Thielemann, M.B.Davies, W.Benz, T.Piran, Astron.Astrophys.341:499-526 (1999)
[2] Luke Bovard, Dirk Martin, Federico Guercilena, Almudena Arcones, Luciano Rezzolla, and Oleg Korobkin, Phys. Rev. D 96, 124005 (2017)
[3] Sho Fujibayashi, Kenta Kiuchi, Nobuya Nishimura, Yuichiro Sekiguchi, Masaru Shibata, The Astrophysical Journal 860:64 (2018)
[4] Andreas Bauswein, Oliver Just, Hans-Thomas Janka, Nikolaos Stergioulas, The Astrophysical Journal Letters, 850:L34 (2017)
[5] B.P. Abbott et al., Phys. Rev. Lett. 119, 161101 (2017)
[6] B.P. Abbott et al., The Astrophysical Journal Letters, 848:L12 (2017)
[7] B. P. Abbott, et al., arXiv:1805.11579 (2018)
[8] B. P. Abbott, et al., arXiv:1805.11581 (2018)
[9] Katerina Chatziioannou, Carl-Johan Haster, Aaron Zimmerman, Phys. Rev. D 97,104036 (2018).
[10] F. J. Fattoyev, J. Piekarewicz, C. J. Horowitz, Phys. Rev. Lett. 120, 172702 (2018)
[11] Sergey Postnikov, Madappa Prakash and James Lattimer, Phys.Rev.D 82, 024016, (2010)
[12] James M. Lattimer and Madappa Prakash, Phys. Rep. 621, 127-164 (2016).
[13] L. Lindblom, Astrophys. J. 398, 569 (1992).
[14] Bao-An Li, Lie-Wen Chen, Che Ming Ko, Phys. Rep. 464, 113-281 (2008).
[15] Pawel Danielewicz, Pardeep Singh, Jenny Lee, Nucl. Phys. A 958, 147-186 (2017)
[16] M. Dutra, O. Lourenco, J.S. Sa Martins, A. Delfino, J.R. Stone, P.D. Stevenson, Phys. Rev. C 85, 035201 (2012)
[17] F. J. Fattoyev, J. Carvajal, W. G. Newton, Bao-An Li, Phys. Rev. C 87, 015806 (2013)
[18] Gordon Baym, Christopher Pethick, Peter Sutherland, Astrophys J., 170:299-137 (1971)
[19] C. P. Lorenz, D. G. Ravenhall, and C. J. Pethick, Phys. Rev. Lett. 70, 379 (1993)
[20] A. S. Schneider, C. J. Horowitz, J. Hughto, D. K. Berry, Phys. Rev. C88, 065807 (2013)
[21] A. S. Schneider, D. K. Berry, C. M. Briggs, M. E. Caplan, C. J. Horowitz, Phys. Rev. C90, 055805 (2014)
[22] D. K. Berry, M. E. Caplan, C. J. Horowitz, G. Huber, A. S. Schneider, Phys. Rev. C94, 055801 (2016)
[23] F. J. Fattoyev, C. J. Horowitz, B. Schuetrumpf, Phys. Rev. C 95, 055804 (2017)
[24] J.W. Negele, D. Vautherin, Nuclear Physics A 207:298-320 (1973)
[25] Bennett Link, Richard I. Epstein, James M Lattimer, Phys. Rev. Lett., 83, 3362-3365 (1999)
[26] J. Carriere, C.J. Horowitz, J. Piekarewicz, Astrophys J, 593:464-471 (2003)
[27] E. Annala, T. Gorda, A. Kurkela, A. Vuorinen, Phys. Rev. Lett. 120, 172703 (2018)
[28] J.M. Lattimer, M. Prakash, Science 304, 536 (2004)
[29] W.G. Lynch, M.B. Tsang, arXiv:1805.10757
[30] B.A. Brown, Phys. Rev. Lett. 111, 232502 (2013).
[31] M. Kortelainen, T. Lesinski, J. Moré, W. Nazarewicz, J. Sarich, N. Schunck, M. V. Stoitsov, and S. Wild, Phys. Rev. C 82, 024313 (2010).
[32] D.H. Youngblood, H.L. Clark, Y.W. Lui, Phys. Rev. Lett. 82, 691 (1999)
[33] A. Tamii, et al., Phys. Rev. Lett. 107, 062502 (2011).
[34] Zhen Zhang and Lie-Wen Chen, Phys. Rev. C 92, 031301(R) (2015).
[35] M.B. Tsang, Y.X. Zhang, P. Danielewicz, M. Famiano, Z.X. Li, W.G. Lynch and A.W. Steiner, Phys. Rev. Lett. 102, 122701 (2009).
[36] D.D.S. Coupland, M. Youngs, Z. Chajecki, W.G. Lynch, M. B. Tsang, Y.X. Zhang, M.A. Famiano, T. Ghosh, et al. Phys. Rev. C 94 011601(R) (2016).

[37] Pawel Danielewicz, Roy Lacey, and William G. Lynch, Science 298, 1592 (2002).
[38] Christian Fuchs, Nucl. Phys. 56, 1-103 (2006)
[39] W.G. Lynch, M.B. Tsang, Y. Zhang, P. Danielewicz, M. Famiano, Z. Li, A.W. Steiner; Prog. Part. Nucl. Phys. 62, 427-432 (2009)
[40] A. LeFèvre, Y. Leifels, W. Reisdorf, J. Aichelin, Ch. Hartnack, Nucl. Phys. A 945, 112 (20).
[41] P. Russotto et al. Phys. Rev. C **94**, 034608
[42] M. Prakash, T. L. Ainsworth and J.M. Lattimer, Phys. Rev. Lett. 61, 2518 (1988).
[43] A.W. Steiner, J.M. Lattimer, E.F. Brown, Astrophys. J. Lett. 765, L5 (2013)
[44] Soumi De, Daniel Finstad, James M. Lattimer, Duncan A. Brown, Edo Berger, and Christopher M. Biwer, (2018), arXiv:1804.08583
[45] I. Tews, J. Margueron, S. Reddy (2018), arXiv:1804.02783
[46] Elias R. Most, Lukas R. Weih, Luciano Rezzolla, Jurgen Schaffner-Bielich, Phys. Rev. Lett. 120, 261103 (2018).
[47] Nai-Bo Zhang, Bao-An Li, Jun Xu, The Astrophysical Journal, 859:90 (2018)
[48] D. Radice, A. Perego, F. Zappa, and S. Bernuzzi, Astrophys. J. Lett. 852, L29 (2018)
[49] Yeunhwan Lim, and Jeremy W. Holt, arXiv: 1803.02803
[50] S. Gandolfi, J. Carlson, S. Reddy, Phys. Rev. C 85, 032801(R) (2012)
[51] J.M. Lattimer, M. Prakash, Astrophys. J. 550, 426 (2001)
[52] P. Danielewicz, Phys. Rev. C 51, 716 (1995)
[53] Jun Xu et al., Phys. Rev. C 93, 044609 (2016)
[54] Ying-Xun Zhang, et. al., Phys. Rev. C 97, 034625 (2018)
[55] M.B. Tsang, J. Estee, H. Setiawan, W.G. Lynch, J. Barney, M.B. Chen, G. Cerizza, P. Danielewicz, J. Hong, P. Morfouace, et. al., Phys. Rev. C 95, 044614 (2017).
[56] R. Shane et al. Nucl. Instrum. Methods A 784, 513 (2015)